\documentclass{article}
\usepackage{amssymb}
\usepackage{amsmath,bm}
\numberwithin{equation}{section}

\usepackage{graphicx}

\newtheorem{theorem}{Theorem}[section]

\newtheorem{remark}[theorem]{Remark}

\newcommand{\D}{\mathrm{d}}

\newcommand{\uu}{{\bf{u}}}

\newcommand{\pp}[2]{ \frac{\partial #1}{\partial #2} }
\begin{document}
\title{ On solutions of the Euler equation for incoherent fluid on a rotating sphere}
\author{
B. G. Konopelchenko $^{1}$  and G.Ortenzi $^{2}$ 
\footnote{Corresponding author. E-mail: giovanni.ortenzi@unito.it }\\
$^{1}$ {\footnotesize INFN, Sezione di Lecce, via Provinciale per Arnesano, 73100  Lecce, Italy } \\
 $^2$ {\footnotesize  Dipartimento di Matematica ``G. Peano'', 
Universit\`{a} di Torino, via Carlo Alberto 10, 10123, Torino, Italy}\\
$^2${\footnotesize   INFN, Sezione di Torino, via Pietro Giuria 1, 10125 Torino, Italy}
} 
\maketitle
\begin{flushright}{\footnotesize {\it In memory of Francesco Calogero \qquad} } \end{flushright}
\abstract{ The motion of compressible, inviscid fluid under the constant pressure on a 
rotating sphere is studied. The hodograph equations for the corresponding Euler equation are presented. They provide us with the class of solutions of the Euler equation
parameterized by two arbitrary functions of two variables. Several particular explicit solutions are given. The blow-up curves, on which the derivatives of velocitiy blows up,
 are described. The limiting cases of slowly and rapidly rotating sphere are considered. The equation describing the deformations of elliptic functions modulus is presented. 
}
\section{Introduction}
\label{sec-intro}
Euler equation on a  rotating two-dimensional sphere is the principal ingredient of a number of models describing atmospheric and ocaenographic flows on 
Earth (see e.g. \cite{Eck60,Ped79,Gill82,L-VI,Sal88,AK98,Val06,Zei07,GZMvH08,MCC20}). Specific approximations and simplifications characterizing these models provide us with various types of results of different 
degrees of applicability and correspondence to real problems (see e.g. \cite{CM88,BMN96,Gat04,CM13,TDBZ14}). However, in spite of numbers of results in this field , the construction of
exact solutions of the Euler equation on a rotating sphere still remains an important task.

In the present paper we consider the most simplified model describing the motion of the incoherent (compressible, inviscid, constant pressure) fluid on the
2-sphere of unit radius $S^2$ rotating at the speed $\omega$ about polar axis. With  the standard metric on $S^2$ the corresponding Euler equation is of the form
(see e.g. \cite{CM88,BMN96,Gat04,CM13,TDBZ14,CG22} )
\begin{equation}
\begin{split}
& \pp{u}{t}+ u \pp{u}{\theta}+v \pp{u}{\phi}=\mathcal{F}_1=\sin(\theta) \cos(\theta) \left( v+\omega\right)^2\, , \\
& \pp{v}{t}+ u \pp{v}{\theta}+v \pp{v}{\phi}=\mathcal{F}_2=-2 \frac{\cos(\theta)}{\sin(\theta)} u \left( v+ \omega \right)\, , \\
\end{split}
\label{EE-Alisei}
\end{equation}
where $0 \leq  \theta< \pi$, $0 \leq  \phi< 2\pi$ are the standard angles on $S^2$, $u=\D \theta/\D t$ and $v=\D \phi/\D t$ are the corresponding velocities and the force $\mathcal{F}^i$,  $i=1,2$ contains both Coriolis and centrifugal contributions. 

We present here the hodograph equations which provide us with exact solutions of the system (\ref{EE-Alisei}). 
The construction is based on the concept 
of integral hypersurface and the method of characteristics. The use of four functionally independent integrals of equations 
of characteristics gives us the class of  the exact generic solutions of the Euler equation (\ref{EE-Alisei}) parameterized by two arbitrary 
functions of two variables. Particular subclasses of solutions, including exact explicit solutions and  solutions periodic in time are presented too. 
The close interrelation of  the solutions of the system (\ref{EE-Alisei}) and those at $\omega=0$ is noted. 

We consider also the particular cases with different interrelations between $v$ and $\omega$.  
The case $\omega=0$ has been considered in the paper \cite{KO25-curved}. the situation with small $\omega$, i.e.  $\frac{\omega}{v}\ll 1$, when one 
approximates $(v+\omega)^2$ by $v^2+2v\omega$ in $\mathcal{F}_1$ corresponds to the case when only the Coriolis force contributes in the r.h.s of (\ref{EE-Alisei}).
 The opposite situation of large $\omega$,  i.e.  $\frac{v}{\omega} \ll1$, when one approximates the force $\mathcal{F}_i$, $i=1,2$
 by $\mathcal{F}_1=\omega^2 \sin(\theta)\cos(\theta)$, $\mathcal{F}_2=-2 \omega \frac{\cos(\theta)}{\sin(\theta)}u$ is analyzed too. Note that it is not the case of the presence of only centrifugal force in the r.h.s. of (\ref{EE-Alisei}).
 
 The reduction $v+\omega=0$ both in generic and only Coriolis force cases are discussed. Interpretation of the corresponding equations as those
 describing special classes of deformation of elliptic function modulus is presented.
 
 Finally, we consider also the solutions of Euler equations written in terms of physical velocities $\tilde{u}=u_1$, $\tilde{v}=v\sin(\theta)$
 

The paper is organized as follows. In section \ref{sec-EEgen} equations of characteristics for the system (\ref{EE-Alisei}) and associated integrals are 
presented. In section \ref{sec-genhodo} hodograph equation are derived and corresponding solutions are discussed. The case of slowly rotating sphere is 
considered in section \ref{sec-slowCor} while the opposite situation of rapidly rotating sphere is discussed in section \ref{sec-bigrot}. Solutions of the
Euler equation for rapidly rotating sphere in presence of only Coriolis forse are discussed in section \ref{sec-bigrotCor}.  The reduction $v+\omega=0$
and equations for the elliptic modulus are considered in section \ref{sec-intvel}. In section \ref{sec-physvel} the hodograph equations for the Euler equations 
at physical velocities are presented. Some particular solutions of the equation (\ref{EE-Alisei}) are discussed in Appendix \ref{app-constantL3}.
Lagrangian structure of system (\ref{chareqn-Cor}) is briefly discussed in the Appendix \ref{app-eqstart}.


\section{Integral hypersurface, characteristics and integrals}
\label{sec-EEgen}
In this section we will analyze the  structure of the characteristics for and incoherent fluid in a rotating sphere described by the equation (\ref{EE-Alisei}). 
\subsection{Integral hypersurface}
Integral hypersurface is one of the central objects in the study of quasilinear PDEs (\cite{CH1,CH2}). 
In our case it is three-dimensional hypersurface in the 
5-dimensional space with coordinates $(t,\theta,\phi,u,v)$ defined by the system of equations
\begin{equation}
S_i(t,\theta,\phi,u,v)=0\, , \qquad i=1,2
\label{genfront}
\end{equation}
such that the resolution of this system with respect to $u$ and $v$ provides us with the solutions of the system (\ref{EE-Alisei}). The functions 
 $S_i(t,\theta,\phi,u,v)$ obey a certain system of linear equations \cite{CH2,Whi}. For the Euler equation (\ref{EE-Alisei}) $S_i$,  $i=1,2$ 
 are two independent solutions of the linear equation
\begin{equation}
\pp{S_i}{t}+ u \pp{S_i}{\theta}+v \pp{S_i}{\phi}+\mathcal{F}_1(t,\theta,\phi,u,v)  \pp{S_i}{u} 
+\mathcal{F}_2(t,\theta,\phi,u,v)  \pp{S_i}{v}=0\, , \qquad i=1,2\, . 
\label{hodoPDE}
\end{equation}
 Any solution of this system provides us, via (\ref{genfront}), with a local solution of the Euler equation (\ref{EE-Alisei}) under the assumption that the 
 matrix $\left(\pp{S_i}{u}, \pp{S_i}{v}\right)$ is invertible. General solutions of the system (\ref{hodoPDE}) depends on 2 functions of 4 variables. The use of such general solution in (\ref{genfront}) gives us a general solution of equations (\ref{EE-Alisei}).
 
 Method of characteristics is the standard method for constructions of solutions of the linear equation (\ref{hodoPDE}). Characteristics for the system 
 (\ref{hodoPDE}) are defined by the equations
 \begin{equation}
\frac{\D \theta}{\D t}=u\, , \qquad
\frac{\D \phi}{\D t}=v\, , \qquad
\frac{\D u}{\D t}=\mathcal{F}_1=\sin(\theta) \cos(\theta) \left(v+ \omega\right)^2\, , \qquad
\frac{\D v}{\D t}=\mathcal{F}_2=-2u\left(v+\omega \right)\mathrm{cotan}(\theta) \, .
\label{chareqn}
\end{equation}

Solutions of the system (\ref{hodoPDE})  are constants along characteristics, i.e.
\begin{equation}
\frac{\D S_i}{\D \tau}=0\, , \qquad i=1,2\, .
\end{equation}
So, $S_i$ are integrals of equations of characteristics. If $(I_1, \dots, I_m)$ are functionally ndependent integrals of the system (\ref{chareqn}), then the 
 functions
 \begin{equation}
 S_i=\phi_i (I_1, \dots, I_m)\, , \qquad i=1,2
 \label{generic-S}
 \end{equation}
where $\phi_i$ are arbitrary functions, are solutions of the system  (\ref{hodoPDE}).
In this case, using (\ref{genfront}), one gets solutions of the Euler equation depending on $m-2$ arbitrary functions of 2 variables. 
One has generic solution if $m=4$.
\subsection{Integrals of characteristics}
So, our task now is construct integrals for characteristics of equations (\ref{genfront}) or, equivalently, for the dynamical system
\begin{equation}
\begin{split}
\frac{\D^2 \theta}{\D \tau^2}&=\sin(\theta)\cos(\theta) \left( \frac{\D \phi}{\D \tau} +\omega \right)^2 \, , \\
\frac{\D^2 \phi}{\D \tau^2}&=-2 \frac{\cos(\theta)}{\sin(\theta)}\left( \frac{\D \phi}{\D \tau} +\omega \right) \, . 
\end{split}
\label{gencharODE-H}
\end{equation}

First, it is a straightforward check that there are 3 integrals linear in $u$ and $v$:
\begin{equation}
\begin{split}
L_1&= - \sin(\theta)\cos(\theta)\cos(\phi+\omega  \tau)\left(\frac{\D \phi}{\D \tau}+\omega\right)-\sin(\phi+\omega \tau)\frac{\D \theta}{\D \tau}\, ,\\
L_2&= - \sin(\theta)\cos(\theta)\sin(\phi+\omega  \tau )\left(\frac{\D \phi}{\D \tau}+\omega\right)+\cos(\phi+\omega  \tau)\frac{\D \theta}{\D \tau}\, ,\\
L_3&=   \sin^2(\theta)\left(\frac{\D \phi}{\D \tau}+\omega\right)\, .
\end{split}
\label{genangmom}
\end{equation}
where the radius of the sphere is $R=1$.
These integrals are not independent on the sphere $S^2$. Indeed, for all values 
$\theta$ and $\phi$ one has the relation
\begin{equation}
\cos(\phi+\omega  \tau) L_1 + \sin(\phi+\omega  \tau) L_2 + \frac{\cos(\theta)}{\sin(\theta)}L_3=0\, .
\end{equation}
This relation is invariant under the dynamics (\ref{gencharODE-H}). Then it is easy
to check that there is integral quadratic in $u$ and $v$, namely,
\begin{equation}
H=\frac{1}{2} \left(  \left( \frac{\D \theta}{\D \tau} \right)^2 + \sin^2(\theta) \left(\frac{\D \phi}{\D \tau}+\omega\right)^2 \right)\, .
\label{gen-energy}
\end{equation}
In appendix \ref{app-eqstart} we relate this quantity  with the Hamiltonian of the 
ODE system (\ref{gencharODE-H}).
The integral $H$ is closely related with integrals (\ref{genangmom}), namely, one has the relation
\begin{equation}
L_1^2 + L_2^2+L_3^2= 2 H\, .
\end{equation}

However, similarly to the case $\omega=0$ \cite{KO25-curved}, the integral (\ref{gen-energy}) is useful for the construction of other two integrals. 
Indeed, using integrals $L_3$ and $H$, one obtains the equations 
\begin{equation}
\frac{\D \theta}{\D \tau}=\sigma \sqrt{2H} \frac{\sqrt{\sin^2(\theta)-K^2}}{\sin(\theta)}
\end{equation}
where $\sigma=\mathrm{sign}(u(0,\theta_0,\phi_0))$ and $K^2 = \frac{L_3^2}{2H}<1$.

Integrating this equation, one obtains the integral
\begin{equation}
I_1=\tau+ \sigma \frac{1}{ \sqrt{2H}} \tilde{Q}(\theta)\, ,
\label{1st-intt}
\end{equation}
where $\tilde{Q}(\theta) =  \arcsin \left(\frac{\cos(\theta)}{\sqrt{1-K^2}} \right)$.
Next, using (\ref{1st-intt}) resolve with respect to $\cos(\theta)$, one finds a solution of the equation 
\begin{equation}
\frac{\D \phi}{\D \tau} +\omega=\frac{L_3}{\sin^2(\theta)}\, ,
\end{equation}
and, consequently, the integral
\begin{equation}
I_2=\phi+ \omega \tau+ 
\mathrm{arctan}\left( K \tan(\sigma \tilde{Q}(\theta)) \right)
-\mathrm{arctan}\left( K \tan( \sqrt{2H}\tau+\sigma  \tilde{Q}(\theta)) \right) 
\, .
\end{equation}
Thus one has 6 integrals of the system (\ref{gencharODE-H}), and, hence, 6 solutions of the equation (\ref{hodoPDE}). 
They are ($\tau \to t$, $\D \theta/\D \tau \to u$, $\D \phi/\D \tau \to v$)
\begin{equation}
\begin{split}
L_1(t,\theta,\phi,u,v)=& - \sin(\theta)\cos(\theta)\cos(\phi+\omega t)\left(v+\omega\right)-\sin(\phi+\omega t)u\, ,\\
L_2(t,\theta,\phi,u,v)=& - \sin(\theta)\cos(\theta)\sin(\phi+\omega t)\left(v+\omega\right)+\cos(\phi+\omega t)u\, ,\\
L_3(\theta,u,v)=&   \sin^2(\theta)\left(v+\omega\right)\, ,\\
H(\theta,u,v)=&\frac{1}{2} \left(u^2 + \sin^2(\theta) \left(v+\omega\right)^2 \right)\,\\
I_1(t,\theta,u,v)=&t+ \sigma \frac{1}{ \sqrt{2H(\theta,u,v)}} Q(u,v,\theta)\, , \\
I_2(t,\theta,\phi,u,v)=&\phi +\omega t+ \mathrm{arctan}\left( \frac{v+\omega}{u}\sin(\theta)\cos(\theta) \right)-\\
&-\mathrm{arctan} \left(\sigma \frac{\sin^2(\theta) (v+\omega)}{\sqrt{2 H(\theta,u,v)}} \tan \left( \sqrt{2H(\theta,u,v) }t + \sigma Q(u,v,\theta)\right) \right)\, . 
\end{split}
\label{gencc-fluid}
\end{equation}
where $Q(u,v,\theta)=\arcsin\left( \cos(\theta) \sqrt{\frac{u^2 + \sin^2(\theta) \left(v+\omega \right)^2 }{u^2 + \sin^2(\theta) \cos^2(\theta) \left(v+\omega\right)^2 }}\right)$.

It is noted that quantities (\ref{gencc-fluid}) are not integrals of equation (\ref{EE-Alisei}).

\section{Hodograph equations and solutions}
\label{sec-genhodo}
 The construction of the previous section has an hodograph  counterpart which is useful for the study of   
 blowup derivatives for the equation (\ref{EE-Alisei}).  
\subsection{Hodograph equation}
The list (\ref{gencc-fluid}) provides us with the different possible choices of arguments of the functions $\phi_i$ in (\ref{generic-S}) and, consequently 
different hodograph equations. Choosing $S_i=\Phi_i(I_1,I_2,H,L_3)$, $i=1,2$ and resolving equations (\ref{genfront}) with respect to $I_1$, $I_2$, 
one gets the hodograph equations 
\begin{equation}
I_1=\Phi_1(H,L_3)\, , \qquad
I_2=\Phi_2(H,L_3)\, ,
\label{hard-hodo}
\end{equation}
where $\Phi_i$, $i=1,2$ are arbitrary functions. 

Choosing instead $S_i=\Phi_1(L_1,L_2,L_3,I_1)$, $i=1,2$, one avoids the complicated integral $I_2$ and obtains the following hodograph equations
\begin{equation}
\begin{split}
&t+\sigma \frac{1}{\sqrt{u^2+\sin^2(\theta) (v+\omega)^2}} Q(u,v,\theta)=\Phi_1(L_1,L_2)\, , \\
&(v+\omega)\sin^2(\theta)=\Phi_2(L_1,L_2)\, ,
\end{split}
\label{simple-hodo}
\end{equation}
where $\Phi_i$, $i=1,2$ are arbitrary functions. It is the straightforward check that the functions $u(t,\theta,\phi)$ and $v(t,\theta,\phi)$ obeying (\ref{hard-hodo})
or (\ref{simple-hodo}) are solutions of the Euler equation (\ref{EE-Alisei}). Indeed, differentiating (\ref{simple-hodo}) with respect to $t, \theta$ and $\phi$, 
one gets the relation
\begin{equation}
\begin{split}
M \begin{pmatrix}
\pp{u}{t} \\ \, \\ \pp{v}{t} 
\end{pmatrix}=& \begin{pmatrix}
1- \pp{\Phi_1}{t}\\ \, \\ -\pp{\Phi_2}{t} 
\end{pmatrix}\, ,\\
M \begin{pmatrix}
\pp{u}{\theta} \\ \, \\ \pp{v}{\theta} 
\end{pmatrix}=& \begin{pmatrix}
 \pp{I_1}{\theta}-\pp{\Phi_1}{\theta}\\ \, \\ 2 (v+\omega) \sin(\theta) \cos(\theta) -\pp{\Phi_2}{\theta}
\end{pmatrix}\, ,\\
M \begin{pmatrix}
\pp{u}{\phi} \\ \, \\ \pp{v}{\phi} 
\end{pmatrix}=& \begin{pmatrix}
-\pp{\Phi_1}{\phi} \\ \,\\ -\pp{\Phi_2}{\phi} 
\label{hodo-gen}
\end{pmatrix}\, ,
 \end{split}
\end{equation}
where the matrix $M$ is
 \begin{equation}
M = \begin{pmatrix}
\pp{\Phi_1}{u}-\pp{I_1}{u}  & \pp{\Phi_1}{v}-\pp{I_1}{v} \\ \, \\ \pp{\Phi_2}{u}  & \pp{\Phi_2}{v}-\sin^2(\theta)
\end{pmatrix}\, .
\end{equation}
Combining relation (\ref{hodo-gen}), one obtains
\begin{equation}
M \begin{pmatrix}
\frac{\D u}{\D t} -(v+\omega)^2 \sin(\theta)\cos(\theta) \\ \, \\ \frac{\D v}{\D t} +2u(v+\omega) \frac{\cos(\theta) }{ \sin(\theta)}
\end{pmatrix} = \begin{pmatrix}
0\\ \, \\ 0
\end{pmatrix} \, ,
\end{equation}
where $\frac{\D}{\D t}=\pp{}{t}+u\pp{}{\theta}+v\pp{}{\phi}$. 
Hence, if $\det(M) \neq 0$ the functions $u$ and $v$ given by (\ref{simple-hodo}) are solutions of the system (\ref{EE-Alisei}).

Derivatives  $\pp{\uu}{t}$, $\pp{\uu}{\theta}$, and $\pp{\uu}{\phi}$ of these solutions blow-up if $\det(M) = 0$, i.e. on the submanifold defined by the equation
\begin{equation}
\sin^2(\theta) \pp{}{u} (I_1-\Phi_1)     - \pp{}{u}(I_1-\Phi_1) \pp{\Phi_2}{v}+\pp{}{v}(I_1-\Phi_1) \pp{\Phi_2}{u}=0\, . 
\end{equation}

It is noted that  due to the invariance under the shift $\omega \to \omega + 2 \pi$  the hodograph equations 
(\ref{simple-hodo})  provide us with the single-valued solutions of the Euler equation (\ref{EE-Alisei}), 
i.e. $\uu(t,\theta,\phi)=\uu(t,\theta,\phi+2\pi)$. If, instead 
of $I_1$ one choses $I_2$, then the solutions of the corresponding hodograph equations 
\begin{equation}
I_1=\Phi_1(L_1,L_2)\, , \qquad (v+\omega) \sin^2(\theta) = \Phi_2(L_1,L_2)\, ,
\label{hodo-nonsingle}
\end{equation}
are generically not single-valued
\subsection{Solutions}
The simplest solution of the hodograph equations (\ref{simple-hodo}) is obtained for constant $\Phi_1$ and $\Phi_2$. In this case $u$ and $v$ are 
independent on $\phi$ and are given by
\begin{equation}
u^2= W^2(t,\theta)-\frac{\Phi^2_2}{\sin^2(\theta)}\, , \qquad v=\frac{\Phi_2}{\sin^2(\theta)}-\omega\, ,
\label{sol-degphi}
\end{equation}
 where the function $W$ is defined by the equation
 \begin{equation}
 (t-\Phi_1) W + \sigma \arcsin \left( \cos(\theta) \frac{W}{\sqrt{W^2-\Phi_2^2}}\right)=0\, .
 \end{equation}
Solution (\ref{sol-degphi}) blows up at the north and south pole while their derivatives blow up us on the curve given by the equation $\pp{I_1}{u}=0$.

One gets a more complicated solution in explicit form in the case of the functions $\Phi_1$ and $\Phi_2$ linear in 
$L_1$ and $L_2$, i.e.
\begin{equation}
\Phi_i= a_i L_1+b_i L_2\, , \qquad i=1,2\, , \quad a_i,b_i \in \mathbb{R}\, .
\end{equation}
Indeed in such a case the second equation (\ref{simple-hodo}) implies that
\begin{equation}
\begin{split}
&A(\omega+v)=Bu\, , \\
&A= \sin^2(\theta)+a_2 \sin(\theta)\cos(\theta)\cos\left(\phi+\omega t\right) + b_2 \sin(\theta)\cos(\theta)\sin \left(\phi+\omega t\right)  \, , \\
&B=-a_2 \sin \left(\phi+\omega t\right) +b_2 \cos \left(\phi+\omega t\right)\, .
\end{split}
\label{lin-exe-constr1}
\end{equation}
Substitution of (\ref{lin-exe-constr1}) into the first equation of (\ref{simple-hodo}) gives
\begin{equation}
\begin{split}
&\alpha u^2 +t u +\beta=0\, , \\
&\alpha= a_1 \sin(\phi+\omega t) - b_1 \cos(\phi+\omega t)+\frac{B}{A} \sin(\theta)\cos(\theta) 
\left[a_1 \cos(\phi+\omega t) + b_1 \sin(\phi+\omega t)\right]\, , \\
&\beta= \ \frac{\sigma A}{\sqrt{A^2+B^2 \sin^2(\theta)}} 
 \arcsin\left( \cos(\theta) \sqrt{\frac{A^2+B^2 \sin^2(\theta)}{A^2+B^2 \sin^2(\theta)\cos^2(\theta)}}\right)  \, .
\end{split}
\label{lin-exe-constr2}
\end{equation}
Thus, one has a solution of the Euler equation (\ref{EE-Alisei}) of the form
\begin{equation}
\begin{split}
u_\pm= & \frac{ -t \pm \sqrt{t^2-4 \beta \alpha}}{2 \alpha} \, , \\
v_\pm=&-\omega + \frac{B}{A} \frac{ -t \pm \sqrt{t^2-4 \beta \alpha}}{2 \alpha}  \, .
\end{split}
\label{lin-exe-sol}
\end{equation}
where $A,B$ and $\alpha,\beta$ functions are given in (\ref{lin-exe-constr1}) and  (\ref{lin-exe-constr2}) .

For other choices of functions $\Phi_1$ and $\Phi_2$ one obtains solutions $\uu(t,\theta,\phi)$ of a rather complicated form.

An interesting class of solutions corresponds to the choice 
\begin{equation}
S_i=\Phi_i(L_1,L_2,L_3)\, ,\qquad i=1,2\, ,
\end{equation}
where $\Phi_1$ and $\Phi_2$ are arbitrary functions. Resolving the equations (\ref{genfront}) with respect to $L_1$ and $L_2$, one obtains the following
hodograph equations
\begin{equation}
\begin{split}
&(v+\omega)\sin(\theta)\cos(\theta)\cos(\phi+\omega t) + u  \sin(\phi+\omega t) = F_1\left(\xi \right)\, ,\\
&(v+\omega)\sin(\theta)\cos(\theta)\sin(\phi+\omega t) - u  \cos(\phi+\omega t) = F_2\left( \xi \right)\, , \\
& \text{with} \quad  \xi \equiv (v+\omega) \sin^2(\theta)\, ,
\end{split}
\label{sol-redangmom}
\end{equation}
where $F_1$ and $F_2$ are arbitrary functions. Differentiating (\ref{sol-redangmom})  with respect to $t$, $\theta$ and $\phi$ one obtains the relations
\begin{equation}
\begin{split}
M \begin{pmatrix}
\pp{u}{t} \\ \, \\ \pp{v}{t} 
\end{pmatrix}
=&
\omega 
\begin{pmatrix}
(v+\omega)\sin(\theta)\cos(\theta)\sin(\phi+\omega t) - u  \cos(\phi+\omega t)\\ \, \\-(v+\omega)\sin(\theta)\cos(\theta)\cos(\phi+\omega t) - u\sin(\phi+\omega t)
\end{pmatrix} \, ,\\ & \\
M \begin{pmatrix}
\pp{u}{\theta} \\ \, \\ \pp{v}{\theta} 
\end{pmatrix}
=&
\begin{pmatrix}
(v+\omega)\sin(2\theta)F_1' - (v+\omega) \cos(2\theta) \cos(\phi+\omega t) \\ \, \\(v+\omega)\sin(2\theta)F_2' - (v+\omega) \cos(2\theta) \sin(\phi+\omega t) 
\end{pmatrix}\, , \\ & \\
M \begin{pmatrix}
\pp{u}{\phi} \\ \, \\ \pp{v}{\phi} 
\end{pmatrix}
=&
\begin{pmatrix}
(v+\omega)\sin(\theta)\cos(\theta)\sin(\phi+\omega t) - u  \cos(\phi+\omega t)\\ \, \\-(v+\omega)\sin(\theta)\cos(\theta)\cos(\phi+\omega t) - u\sin(\phi+\omega t)
\end{pmatrix}\, , \\
 \end{split}
 \label{derhodo-part}
\end{equation}
where 
\begin{equation}
M= \begin{pmatrix} 
\sin(\phi+\omega t) & \sin(\theta)\cos(\theta) \cos(\phi+\omega t)  -F_1' \sin^2(\theta) \\ 
\, \\
-\cos(\phi+\omega t) & \sin(\theta)\cos(\theta) \sin(\phi+\omega t)  -F_2' \sin^2(\theta)
\end{pmatrix}
\end{equation}
and $F'$ denotes the derivative of $F$ with respect to the argument.

Combining linearly the relations (\ref{derhodo-part}), one gets
\begin{equation}
M\begin{pmatrix} 
\frac{\D u}{\D t} - (v+\omega)^2 \sin(\theta)\cos(\theta)   \\ 
\, \\
\frac{\D v}{\D t} +2 u(v+\omega) \frac{\cos(\theta)}{\sin(\theta)} 
\end{pmatrix}
=
\begin{pmatrix} 
0 \\ 
\, \\
0 
\end{pmatrix}
\end{equation}
Thus, when $\det(M) \neq 0$ the functions $u(t,\theta,\phi)$ and $v(t,\theta,\phi)$  are solutions of the Euler equations (\ref{EE-Alisei}).
This class of solutions is parameterized by two arbitrary functions of single variables. Such solutions have specific dependence on the variables $t$ 
and $\phi$, namely, the functions $u$ and $v$ depend only on $\theta$ and $\phi+\omega t$. So they are periodic in time with the period $T=\frac{2 \pi}{\omega}$.
At $\omega=0$ they are stationary solutions of the Euler equation on nonrotating sphere (see \cite{KO25-curved}).

The equation $\det M=0$, defines the blowup curve where the derivatives of $\uu$ blow-up. 

Solutions of the hodographic equation  (\ref{sol-redangmom}) can be constructed explicitly. In fact these two equations are equivalent to the following
\begin{equation}
\begin{split}
  (v+\omega)\sin(\theta)\cos(\theta)-F_1\left(\xi  \right)  \cos(\phi+\omega t)- F_2\left(\xi  \right)  \sin(\phi+\omega t)  =& 0 \, ,\\
F_1\left(\xi \right)\sin(\phi+\omega t) -F_2\left(\xi \right)   \cos(\phi+\omega t)  =& u\, . 
\end{split}
\label{sol-redangmom-eq}
\end{equation}
Thus, the problem is reduced to the resolution of the single first equation in (\ref{sol-redangmom-eq}). This can be done explicitly for a wide range of function $F_1$ and $F_2$. 
For linear functions 
 \begin{equation}
F_i= a_i+b_i \xi \, , \qquad a_i,b_i \in \mathbb{R}\, ,\qquad i=1,2 , 
\end{equation}
the solution is given by
\begin{equation}
\begin{split}
u=&a_1 \sin(\phi+\omega t)-a_2 \cos(\phi+\omega t)
+\frac{(a_1 \cos(\phi+\omega t)+a_2 \sin(\phi+\omega t))(b_1 \sin(\phi+\omega t)-b_2 \cos(\phi+\omega t)) \sin(\theta)}
{\cos(\theta)- (b_1 \cos(\phi+\omega t)+b_2 \sin(\phi+\omega t)) \sin(\theta) }\, , \\
v&= -\omega + \frac{a_1 \cos(\phi+\omega t)+a_2 \sin(\phi+\omega t)}{\sin(\theta)\cos(\theta)-\sin^2(\theta)(b_1 \cos(\phi+\omega t)+b_2 \sin(\phi+\omega t))}\, .
\end{split}
\end{equation}

\begin{figure}[h!]
\begin{center}
\includegraphics[width=.4 \textwidth]{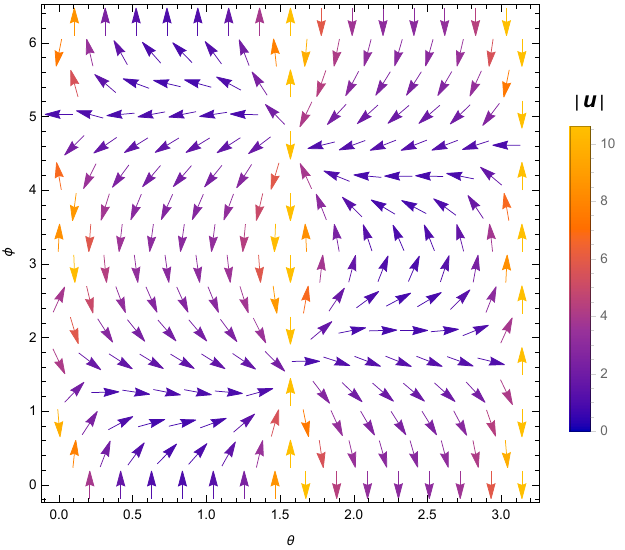}
\caption{The stationary solution (\ref{sol-redangmom-eq}) at $F_1=1$, $F_2=0$ is $\uu=\Big( \sin(\phi+\omega t), 2 \cos(\phi+\omega t)/ \sin(2 \theta) -\omega \Big)$.  It is presented at $\omega=1$, $t=0$.  This solution 
admits singularities for all times at $\theta=0,\pi/2,\pi$. It is locally well defined solution far from the equator and poles.}
\label{Stat-CorandCentr-fig}
\end{center}
\end{figure}

The plot of a simple solution in this family is  in figure \ref{Stat-CorandCentr-fig} .

Subclass of exact solutions of the hodograph equations (\ref{sol-redangmom}) is associated with special functions $F_1$ and $F_2$, namely,
\begin{equation}
F_1=a \xi F\left( \xi \right)\, , \qquad
F_2=b \xi  F\left(\xi \right)\, , \qquad a,b \in \mathbb{R}\, , 
\label{indatprop}
\end{equation}
 and $F$ is an arbitrary function. In this case the solutions of the Euler equation are given by 
 \begin{equation}
 \begin{split}
 u&= -\mathrm{cotan}(\theta)\mathrm{cotan}(\phi+\alpha+\omega t) F^{-1} \left( \sqrt{a^2+b^2} \frac{\mathrm{cotan}(\theta)}{\sin(\phi+\alpha+\omega t)} \right) \, , \\
 v&= -\omega+ \frac{1}{\sin^2(\theta)} F^{-1} \left( \sqrt{a^2+b^2} \frac{\mathrm{cotan}(\theta)}{\sin(\phi+\alpha+\omega t)} \right) \, .
 \end{split}
 \end{equation}
where $\sin(\alpha)\equiv \frac{a}{\sqrt{a^2+b^2}}$ and $F^{-1}(\xi)$ denotes the function inverse  to $F(\xi)$.


With the particular choice  
\begin{equation}
F(\xi)=\sqrt{-\log(\xi)}
\end{equation}
one has ($F^{-1}(\xi)=\exp(-\xi^2)$)
\begin{equation}
\begin{split}
u&= -\mathrm{cotan}(\theta)\mathrm{cotan}(\phi+\alpha+\omega t) \exp\left(- (a^2+b^2) \frac{\mathrm{cotan}^2(\theta)}{\sin^2(\phi+\alpha+\omega t)} \right) \, , \\
 v&= -\omega+ \frac{1}{\sin^2(\theta)} \exp \left( -{(a^2+b^2)} \frac{\mathrm{cotan^2}(\theta)}{\sin^2(\phi+\alpha+\omega t)} \right) \, .
\end{split}
\end{equation}
This solution has the following behavior near the north pole 
\begin{equation}
\begin{split}
u&\sim  -\frac{\mathrm{cotan}(\phi+\alpha+\omega t)}{\theta} \exp\left( - \frac{a^2+b^2}{\theta^2 \sin^2(\phi+\alpha+\omega t)} \right) \, , \\
 v&\sim  \frac{1}{\theta^2}  \exp\left( - \frac{a^2+b^2}{\theta^2 \sin^2(\phi+\alpha+\omega t)} \right)  \, , \qquad \theta \to 0\, ,
\end{split}
\end{equation}
while near the equator it behaves as
\begin{equation}
\begin{split}
u&\sim -\left( \frac{\pi}{2} - \theta \right) \mathrm{cotan}(\phi+\alpha+\omega t) \exp\left(- \frac{a^2+b^2}{ \left( \frac{\pi}{2} - \theta \right)^2 \sin^2(\phi+\alpha+\omega t)} \right) \, , \\
 v&\sim -\omega+ \frac{1}{\sin^2(\theta)} \exp\left(- \frac{a^2+b^2}{ \left( \frac{\pi}{2} - \theta \right)^2 \sin^2(\phi+\alpha+\omega t)} \right) \, ,
  \qquad \theta \to \frac{\pi}{2}\, .
\end{split}
\end{equation}
Note that the behavior of $u$ and $v$ changes periodically in time with period $T=\frac{2 \pi}{\omega}$.
\subsection{Single-valued solutions}
It was already noted that the hodograph equations (\ref{simple-hodo}) provide us with the single-valued  solutions of the Euler equation
(\ref{EE-Alisei}), while those given by the hodograph equation (\ref{hodo-nonsingle}) are not single-valued.

In general, to construct the single-valued solutions of the Euler equation one should to consider the functions $\Phi_i$, $i=1,2$ 
in the expression (\ref{generic-S}) which are invariant under the shift $\phi \to \phi+2\pi$, i.e.  
$\Phi_i(t,\theta,\phi+2\pi,u,v)=\Phi_i(t,\theta,\phi,u,v)$, $i=1,2$ or, equivalently, one should chose the integrals $I_1,\dots,I_4$ which
are invariant under such a  shift.

In our case five integrals from those given by (\ref{gencc-fluid}), namely, $L_1,L_2,L_3,H,I_1$ are invariant while $I_2$ is not. 
So, it is sufficient  to consider the integral $\mathcal{T}(I_2)$ such that $\mathcal{T}(I_2)=\mathcal{T}(I_2+2\pi)$ where $\mathcal{T}(I_2)$
is a generic periodic function. Choosing $S_i$ in the form
\begin{equation}
S_i=\Phi(L_1,L_2,L_3,H,I_1,\mathcal{T}(I_2))\, , \qquad i=1,2\, ,
\end{equation}
one gets the single-valued solutions of the Euler equation (\ref{EE-Alisei}). For example, one can choose  $\mathcal{T}(I_2)=\sin(I_2)$.
\subsection{Relation with the nonrotating case:  solutions periodic in time.}
Comparing the the formulae presented above with those derived in \cite{KO25-curved} for the case $\omega=0$, one readily observes their close interrelation.
In fact, one can construct solutions of the system (\ref{EE-Alisei}) starting from that in the nonrotating case ($\omega=0$), i.e.
\begin{equation}
\begin{split}
& \pp{\tilde{u}}{t}+ \tilde{u} \pp{\tilde{u}}{\theta}+\tilde{v} \pp{\tilde{u}}{\phi}= \tilde{v}^2 \sin(\theta) \cos(\theta) \, , \\
& \pp{\tilde{v}}{t}+ \tilde{u} \pp{\tilde{v}}{\theta}+\tilde{v} \pp{\tilde{v}}{\phi}=-2 \tilde{u} \tilde{v} \mathrm{cotan}(\theta) \, . \\
\end{split}
\label{EE-Sphere}
\end{equation}
It is an easy check that if $\tilde{u}$ and $\tilde{v}$ satisfy (\ref{EE-Sphere}) then  the functions
\begin{equation}
u(t,\theta,\phi)=\tilde{u}(t,\theta,\phi+\omega t)\, , \qquad v(t,\theta,\phi)=\tilde{v}(t,\theta,\phi+\omega t)-\omega\, ,
\label{map-rnr}
\end{equation}
are solutions of the equations (\ref{EE-Alisei}).

It is noted that due to the relation (\ref{map-rnr}) all stationary single valued solutions of the Euler system (\ref{EE-Sphere}), i.e. solutions such that 
$\tilde{u}_t=\tilde{v}_t=0$ with $\tilde{u}(\theta,\phi+2 \pi)=\tilde{u}(\theta,\phi)$ and $\tilde{v}(\theta,\phi+2\pi)=\tilde{v}(\theta,\phi)$,  
become the periodic in time solutions of the Euler equations (\ref{EE-Alisei}) with period $T=\frac{2\pi}{\omega}$.

Of course is quite natural that the transition from the non-rotating sphere to the rotating one involves the changes 
$\phi \to \phi + \omega t$ and $v \to v+ \omega$. However, the existence of the transformation (\ref{map-rnr}) does not trivialize the formulae derived
above without any reference to the non-rotating case. They will be rather useful in the discussion of several particular limits of equation (\ref{EE-Alisei}).

Inverse transformations $\tilde{u} \to u$, $\tilde{v} \to v$ obviously is given by the relation
\begin{equation}
\tilde{u}(t,\theta,\phi)={u}(t,\theta,\phi-\omega\,  t) \, , \qquad \tilde{v}(t,\theta,\phi)={v}(t,\theta,\phi-\omega\,  t) +\omega\, .
\label{invmap-rnr}
\end{equation}
So, the transformations (\ref{map-rnr}), (\ref{invmap-rnr}) represent one-to-one mapping between the solutions
of the Euler equations for rotating and non-rotating cases.

For solutions of the Euler equations (\ref{EE-Alisei}) and (\ref{EE-Sphere}) not depending on $\phi$ the above transformation is simplified  to
the following trivial relation
\begin{equation}
{u}(t,\theta)=\tilde{u}(t,\theta) \, , \qquad {v}(t,\theta)=\tilde{v}(t,\theta) -\omega\, .
\label{invmap-rnr-nophi}
\end{equation}

It is noted the transformation (\ref{invmap-rnr}) is in general different from the formal limit $\omega \to 0$ for the solutions of the equation (\ref{EE-Alisei}).

\section{Slowly rotating sphere: Coriolis force.}
\label{sec-slowCor}
Here we begin to consider  particular limits for the Euler equation (\ref{EE-Alisei}), viewing the rotation speed  $\omega$ as a parameter.

The first natural limit corresponds to small $\omega$, i.e.   $\frac{\omega}{v} \ll 1$  for all $t,\theta,\phi$. In this case it is natural to drop the $\omega^2$
term in the r.h.s.  of (\ref{EE-Alisei}) and to leave only terms linear in $\omega$. The corresponding system is
\begin{equation}
\begin{split}
& \pp{u}{t}+ u \pp{u}{\theta}+v \pp{u}{\phi}=\sin(\theta) \cos(\theta) v \left( v+2\omega\right)\, , \\
& \pp{v}{t}+ u \pp{v}{\theta}+v \pp{v}{\phi}=-2 \mathrm{cotan}{(\theta)} u \left( v+ \omega \right)\, , \\
\end{split}
\label{EE-smallrot}
\end{equation}
This system represents the Euler equation for the  the incoherent fluid on the rotating sphere in the situation when one neglects the contribution
of the centrifugal force and only effects of the Coriolis force remains.  Such a situation has been discussed in several papers (see e.g. \cite{Gat04})

Here we will construct hodograph equations for the system (\ref{EE-smallrot}) and discuss properties of its solutions.

\subsection{Characteristics and integrals}

We will proceed in the same way as in the general case.
According to the general hodograph equations approach one needs first to find generic integral hypersurface given by the system of two equations
\begin{equation}
S_i(t,\theta,\phi,u,v)=0\, , \qquad i=1,2
\end{equation}
where $S_i$,  $i=1,2$ are two independent solutions of the linear equation
\begin{equation}
\pp{S_i}{t}+ u \pp{S_i}{\theta}+v \pp{S_i}{\phi}+\sin(\theta) \cos(\theta) \left( v^2+2 \omega v\right)  \pp{S_i}{u} 
-2 \frac{\cos(\theta)}{\sin(\theta)} \left( uv+2 \omega u\right)  \pp{S_i}{v}=0\, , \qquad i=1,2\, . 
\label{hodoPDE-Cor}
\end{equation}
Solutions of the linear equation (\ref{hodoPDE-Cor}) can be found by the method of characteristics. Equations of characteristics for the equation (\ref{hodoPDE-Cor})
are of the form
\begin{equation}
\frac{\D \theta}{\D  \tau}=u\, , \qquad
\frac{\D \phi}{\D  \tau}=v\, , \qquad
\frac{\D u}{\D  \tau}=\sin(\theta) \cos(\theta) \left(v^2+2\omega v\right)\, , \qquad
\frac{\D v}{\D  \tau}=-2\frac{\cos(\theta)}{\sin(\theta)} \left(uv+\omega u\right)\, .
\label{chareqn-Cor}
\end{equation}
 Equivalently
 \begin{equation}
 \begin{split}
 \frac{\D^2 \theta}{ \D  \tau^2}&=\sin \theta \cos \theta \left( \left( \frac{\D \phi}{\D  \tau}+\omega\right)^2-\omega^2\right)\qquad \, ,\\
 \frac{\D^2 \phi}{ \D  \tau^2}&=-2 \mathrm{cotan}(\theta)\frac{\D \theta}{\D  \tau} \left( \frac{\D \phi}{\D  \tau}+\omega\right) \qquad \, .
 \end{split}
 \label{char-Newton}
 \end{equation}
There is only one integral of the dynamical system  (\ref{chareqn-Cor}) linear in velocity, it is
\begin{equation}
L_3= \sin^2(\theta) \left( \frac{\D \phi}{\D  \tau}+\omega\right)\, ,
\label{mom-Cor}
\end{equation}
while an integral quadratic in velocities is
 \begin{equation}
H= \frac{1}{2} \left(\left( \frac{\D \theta}{\D  \tau}\right)^2+  \sin^2(\theta) \left( \frac{\D \phi}{\D  \tau} \right)^2  \right)\, .\\
\label{En-Cor}
\end{equation}
 Note that the integral (\ref{En-Cor}) is the integral (\ref{gen-energy}) at $\omega=0$  while the integral (\ref{mom-Cor}) retain a dependence on $\omega$.
 
In order to find other integrals one has to solve equations (\ref{chareqn-Cor}). 
Using (\ref{mom-Cor}) and (\ref{En-Cor}),  one gets
\begin{equation}
\left( \frac{\D \theta}{\D  \tau} \right)^2=2H +2 \omega L_3 -\frac{L_3^2}{\sin^2 \theta}-\omega^2\sin^2( \theta)\, .
\label{Weierred-Cor}
\end{equation}
Separating the variables in (\ref{Weierred-Cor}), one obtains  
\begin{equation}
\D  \tau=\pm \frac{\sin(\theta) \D \theta}{\sqrt{(2H+2 \omega L_3) \sin^2(\theta)-L_3^2-\omega^2\sin^4(\theta)}}\, .
\label{int-char-1D}
\end{equation}
In terms of the variable  $x=\cos(\theta)$ the equation (\ref{int-char-1D}) assumes the form
\begin{equation}
\D  \tau=\mp \frac{ \D x}{(\sqrt{c+b x^2 +a x^4 }}
\label{int-char-1D-cv}
\end{equation}
with and $a=-\omega^2$, $b=2 \omega^2-2H-2\omega L_3$, and $c=2H+2 \omega L_3-L_3^2-\omega^2$.

It is well known that the integration of the l.h.s of the relation (\ref{int-char-1D-cv}) is expressed in terms of the incomplete elliptic integral of the first kind
(see e.g. \cite{BF71}). 
%
%
We will present the result of integration in the form adapted to our purpose. First, one rewrite (\ref{int-char-1D-cv}) in the form 
\begin{equation}
\omega \D  \tau = \pm \frac{\D x}{\sqrt{(A_+ -x^2)(x^2-A_-)}} \, ,
\label{int-char-1D-zero}
\end{equation}
where $A_\pm$ are roots of the equation $-\omega^2 y^2+by+c=0$, i.e.
\begin{equation}
A_\pm = \frac{\omega^2-(H+\omega L_3) \pm \sqrt{ {(H+\omega L_3)^2}- L_3^2 \omega^2 }}{\omega^2}  
=1- \frac{H+\omega L_3}{\omega^2} \left( 1 \mp  \sqrt{1-\frac{\omega^2 L_3^2}{(H+\omega L_3)^2}}\right) 
\, .
\end{equation}
From the last expression it is obvious that $A\pm < 1$ and ${(H+\omega L_3)^2}- L_3^2 \omega^2=H (H+2 \omega L_3)>0$ for all possible initial data since
\begin{equation}
H>0\, , \qquad  H+2 \omega L_3 =
 \frac{1}{2} \left(\left( \frac{\D \theta}{\D  \tau}\right)^2+  \sin^2(\theta) \left( \frac{\D \phi}{\D  \tau} +2\omega \right)^2  \right)>0\, .
\end{equation}
Then the real valuedness of (\ref{int-char-1D-zero}) requires that $A_+>0$.

Under these conditions the integration of (\ref{int-char-1D-zero}) gives (see e.g. \cite{BF71}, formula 217.00) 
\begin{equation}
-\frac{1}{\sqrt{A_+}} F \left(\arcsin \left( \sqrt{\frac{A_+ -x^2}{A_+ - A_-}} \right) ,  \sqrt{\frac{A_+ -A_-}{A_+ }} \right) = \pm \omega  \tau + const\, ,
\end{equation}
where $F(\xi,k)$ is the incomplete elliptic integral of first king with modulus $k$. So, for the equation of characteristics (\ref{chareqn-Cor}) one has the integral
\begin{equation}
{P}= \pm \sqrt{A_+} \omega  \tau + F \left(\arcsin \left( \sqrt{\frac{A_+ -x^2}{A_+ - A_-}} \right) ,  \sqrt{\frac{A_+ -A_-}{A_+ }} \right)\, .
\label{cqchar-enCor}
\end{equation}
 In order to find the fourth integral one has to integrate the equation (\ref{mom-Cor}), i.e. the equation
 \begin{equation}
 \D \phi = \frac{L_3}{1-x^2(\tau)} \D t-\omega \D \tau
 \label{ODEmom-Cor}
 \end{equation}
where $x(t)$ is given by (\ref{cqchar-enCor}). The relation (\ref{cqchar-enCor}) implies that
\begin{equation}
 \sqrt{\frac{A_+ -x^2}{A_+ - A_-}}=\mathrm{sn} \left( \pm \sqrt{A_+} \omega \tau +const, k \right)\, ,
\end{equation}
where $\mathrm{sn} (F(\arcsin(\xi),k))=\xi$ and $k=\sqrt{\frac{A_+ -A_-}{A_+ }} $.  So
\begin{equation}
x^2(\tau)=A_+ - (A_+ - A_-) \mathrm{sn}^2 \left( \pm \sqrt{A_+} \omega \tau +const, k \right)\, .
\end{equation}
Hence the relation (\ref{ODEmom-Cor}) becomes
\begin{equation}
\D \phi= \frac{L_3 \D \tau}{1-A_+  + (A_+ - A_-) \mathrm{sn}^2 \left( \pm \sqrt{A_+} \omega \tau +const, k \right)}-\omega \D \tau\, ,
\end{equation}
or equivalently
\begin{equation}
\D \phi= \frac{L_3}{\sqrt{A_+ \omega}(1-A_+) }\frac{ \D y}{1- \alpha^2 \mathrm{sn}^2 \left( y, k \right)} -\omega \D \tau\, ,
\label{intmom-Cor}
\end{equation}
where $y= \sqrt{A_+} \omega \tau+const$ and $\alpha^2=\frac{A_+ - A_-}{A_+ - 1}$, $k=\sqrt{\frac{A_+ -A_-}{A_+ }} $ 
(cfr. notations in \cite{BF71} formula (110.04) ).
Since
\begin{equation}
\int \frac{ \D y}{1- \alpha^2 \mathrm{sn}^2 \left( y, k \right)}=\Pi(y,\alpha^2,k)\, ,
\end{equation}
where $\Pi(y,\alpha^2,k)$ is the incomplete elliptic integral of third kind \cite{BF71}. The integration of (\ref{intmom-Cor}) gives
\begin{equation}
N=\phi+\omega \tau- \frac{L_3}{\omega (1-A_+) \sqrt{A_+ } } \Pi \left( \omega  \sqrt{A_+ } t, \frac{A_+ - A_-}{A_+ - 1},\sqrt{\frac{A_+ -A_-}{A_+ }}\right)=const.\, .
\label{cqchar-momCor}
\end{equation}
which is the fourth integral.  

Particular forms of $\Pi(y,\alpha^2,k)$ for different values of $\alpha^2$ in the interval  $-\theta <\alpha^2<-\infty$ are given in
\cite{BF71}, formulae (431.01)-(436.01).

${P}$ and $N$  given in (\ref{cqchar-enCor}) and (\ref{cqchar-momCor}) ,as can be checked by a direct computation, are integrals of equations 
(\ref{chareqn-Cor}). 

\subsection{Hodograph equation and solutions}
The quantities ${P}$ and $N$ constructed above provide us, together  with $H$ and $L_3$, four functionally independent integrals of characteristics
(\ref{chareqn-Cor}).  It is a straightforward check that formulae (\ref{mom-Cor}), (\ref{En-Cor}), (\ref{cqchar-enCor}) and (\ref{cqchar-momCor})
with the substitution
\begin{equation}
\begin{split}
H=& \frac{1}{2}u^2+\frac{1}{2}v^2\sin^2(\theta)\, ,\\
L_3=& (v+\omega) \sin^2(\theta)\, ,\\
A_\pm=& \frac{1}{2\omega^2} \left\{2 \omega^2-
\left[u^2+\sin^2(\theta) ((v+\omega)^2+\omega^2)
\right]
\pm \sqrt{2H u^2+\sin^2(\theta) (v+2\omega)^2}
\right\}
\end{split}
\end{equation}
 represent solutions of equation (\ref{hodoPDE-Cor}) with $\tau \to t$.

So one gets the following general solution for the functions $S_i$,
\begin{equation}
S_i=\Phi_1(H,L_3,M,N)\, .
\end{equation}
Resolving the equation $S_i=0$, $i=1,2$ w.r.t. $M$ and $N$, one obtains the hodograph equations 
\begin{equation}
{P}=\Psi_1(H,L_3)\, , \qquad N=\Psi_2(H,L_3)\, ,  
\label{hodosol-Cor}
\end{equation}
where $\Psi_1$ and $\Psi_2$ are arbitrary functions.

The check that solutions of the hodograph equation  (\ref{hodosol-Cor}) gives us solutions of the Euler equation (\ref{EE-smallrot})is ismilar to that given in section \ref{sec-genhodo}, Indeed, differentiating (\ref{hodosol-Cor}) with respect to $t$, $\theta$ and $\phi$, one obtains
\begin{equation}
M 
\begin{pmatrix}
\pp{u}{t} \\ \, \\ \pp{v}{t} 
\end{pmatrix} = 
\begin{pmatrix}
\pp{{P}}{t} \\ \, \\ \pp{N}{t} 
\end{pmatrix} \, , \qquad
M 
\begin{pmatrix}
\pp{u}{\theta} \\ \, \\ \pp{v}{\theta} 
\end{pmatrix} = 
\begin{pmatrix}
\pp{{P}}{\theta} \\ \, \\ \pp{N}{\theta} 
\end{pmatrix} \, , \qquad
M 
\begin{pmatrix}
\pp{u}{\phi} \\ \, \\ \pp{v}{\phi} 
\end{pmatrix} = 
\begin{pmatrix}
\pp{{P}}{\phi} \\ \, \\ \pp{N}{\phi} 
\end{pmatrix} \, ,
\end{equation}
where
\begin{equation}
M=
\begin{pmatrix}
u \pp{\Psi_1}{u}- \pp{{P}}{u} & v \sin^2(\theta) \pp{\Psi_1}{H}+\sin^2(\theta) \pp{\Psi_1}{L_3}-\pp{{P}}{v}
\\ \, \\
u \pp{\Psi_2}{u}- \pp{N}{u} & v \sin^2(\theta) \pp{\Psi_2}{H}+\sin^2(\theta) \pp{\Psi_2}{L_3}-\pp{N}{v}
\end{pmatrix}\, .
\end{equation}
and ${P}={P}(t,\theta,\phi,u,v)$, $N=N(t,\theta,\phi,u,v)$.
Combining the previous relations, one gets
\begin{equation}
M
\begin{pmatrix}
\frac{\D u}{\D t} -\sin(\theta) \cos(\theta) \left( v^2+2v\omega\right)
\\ \, \\
\frac{\D u}{\D t} + 2 \mathrm{cotan}{(\theta)} u \left( v+ \omega \right)
\end{pmatrix}
=
\begin{pmatrix}
0
\\ \, \\
0
\end{pmatrix}\, .
\end{equation}
So, if $\det(M) \neq 0$ the functions $u$ and $v$ obey the equation (\ref{EE-smallrot}).
If, instead, $\det(M) = 0$ the derivatives of $u$ and $v$ blow up.  The equation $\det(M) = 0$ defines the blow-up surface.

Finally, we note that the integrals ${P}$ and $N$ can be rewritten in different equivalent forms. In particular, performing in (\ref{cqchar-enCor}) and
(\ref{cqchar-momCor}) the reciprocal modulus transformation $k F(\phi, k)=F(\phi,1/k)$, $\sin(\psi)=k\sin{\phi}$ (see \cite{BF71}) one gets the
integrals 
\begin{equation}
\begin{split}
{P}^*=& \pm \sqrt{A_+} \omega t 
+ \sqrt{\frac{A_+}{A_+ - A_-}}F \left(\arcsin \left( \sqrt{\frac{A_+ -\cos^2(\theta)}{A_+ }} \right) ,  \sqrt{\frac{A_+ }{A_+ -A_-}} \right)\, .\, , \\
N^*=& \phi+\omega t- \frac{L_3 \sqrt{A_+ - A_-}}{\omega (1-A_+)  } 
\Pi \left( \arcsin \left( \sqrt{\frac{A_+ - A_-}{A_+}} \sin(\sqrt{A_+} \omega t)\right), \frac{A_+}{A_+ - 1},\sqrt{\frac{A_+ }{A_+}}\right) \, .
\end{split}
\end{equation}
Hodograph equations 
\begin{equation}
{P}^*=\Psi_1^* (H,L_3)\, , \qquad N^*=\Psi_2^* (H,L_3)\, , 
\end{equation}
provide us with the same class of solutions as those (\ref{hodosol-Cor}).

Simple-valued solutions of the equation (\ref{EE-smallrot}) are provided by the hodograph equations associated with the functions
$S_i$, $i=1,2$ of the form $S_i=\Phi_i(H,L_3,{P},\mathcal{T}(N))$, $i=1,2$, i.e.
\begin{equation}
{P}=\Psi_1(H,L_3)\, , \qquad \mathcal{T}(N)=\Psi_2(H,L_3)\, ,
\end{equation}
where $ \mathcal{T}(N)$ is an arbitrary function such that $ \mathcal{T}(N+2\pi)= \mathcal{T}(N)$.

\subsection{Special classes of solutions and constraints}
Choosing different forms of the functions $S_i$, one obtains particular subclasses of solutions. For instance, in the case 
$S_i=\Phi_i(H,L_3,{P})$, $i=1,2$
the solution of the corresponding hodograph equation depend explicitly only on $t$ and $\theta$. 

Considering one function $S$ instead of two, one gets a constraint between $u$ and $v$. For example, let us impose the relation $S(H,L)=0$
where $S$ is some function. Resolving this relation, one obtains $H=\Phi(L_3)$  where $\Phi$ is some function, i.e.
\begin{equation}
u^2=-v^2\sin^2(\theta)+2\Phi((v+\omega)\sin^2(\theta))\, .
\label{exe-constr-Cor}
\end{equation}
It is a direct check that the constraint (\ref{exe-constr-Cor}) is admissible for the system (\ref{EE-smallrot}) which is reduced to the single equation
\begin{equation}
\pp{v}{t}\pm \sqrt{-v^2\sin^2(\theta)+2\Phi((v+\omega)\sin^2(\theta))} \pp{v}{\theta}+v \pp{v}{\phi}
=\mp \mathrm{cotan}{(\theta)}  \left( v+ \omega \right)  \sqrt{-v^2\sin^2(\theta)+2\Phi((v+\omega)\sin^2(\theta))}  \, .
\end{equation}

\section{Rapidly rotating sphere}
\label{sec-bigrot}
In the case of high speed $\omega$, i.e. when $\omega/|v| \gg 1$ the force in the r.h.s. of the Euler equation (\ref{EE-Alisei}) is dominated by 
the highest terms in $\omega$.  So, the Euler equation 
assumes the form
\begin{equation}
\begin{split}
& \pp{u}{t}+ u \pp{u}{\theta}+v \pp{u}{\phi}=\sin(\theta) \cos(\theta)\, \omega^2\, , \\
& \pp{v}{t}+ u \pp{v}{\theta}+v \pp{v}{\phi}=-2  u  \mathrm{\, cotan}(\theta)\,  \omega\, . \\
\end{split}
\label{EE-bigrot}
\end{equation}
Equations for characteristics are
\begin{equation}
\frac{\D \theta}{\D \tau}=u\, , \qquad
\frac{\D \phi}{\D \tau}=v\, , \qquad
\frac{\D u}{\D \tau}=\omega^2\sin(\theta) \cos(\theta) \, , \qquad
\frac{\D v}{\D \tau}=-2 \omega \mathrm{cotan}(\theta) \, .
\label{chareqn-bigrot}
\end{equation}
Equations (\ref{chareqn-bigrot}) immediately provides us with two integrals
\begin{equation}
I_1=\frac{1}{2}u^2-\frac{1}{2} \omega^2 \sin^2(\theta)\, , \qquad I_2=v+2 \omega \log\sin(\theta)\, .
\label{enmom-bigrot}
\end{equation}

These integrals are, in fact, certain limits of two particular elements of the ring of integrals generated by $H$  (\ref{gen-energy}) and $L_3$ (\ref{genangmom}) in the general system. Indeed the integral
\begin{equation}
I^*_1=H-\omega L_3=\frac{1}{2}u^2-\frac{1}{2}\omega^2\left( 1- \left(\frac{v}{\omega}\right)^2\right)\sin^2(\theta)\, ,
\end{equation}
admits as limit of big $\omega$ 
\begin{equation}
\lim_{\frac{v}{\omega}\to 0}I^*_1 =I_1\, .
\end{equation}
In order  to obtain the integral $I_2$ one starts with the integral 
\begin{equation}
I^*_2=\omega \log \frac{L_3}{\omega}.
\end{equation}
In the limit    $v/\omega \to 0$ one has
\begin{equation}
\lim_{\frac{v}{\omega}\to 0}I^*_2=\lim_{\frac{v}{\omega}\to 0}  \omega \log \left[ \left(1+\frac{v}{\omega}\right)\sin^2(\theta) \right] =
\lim_{\frac{v}{\omega}\to 0}    \left\{\omega  \log \left[ \left(1+\frac{v}{\omega}\right)\right]+\omega \log \sin^2(\theta)\right\}=I_2\, .
\end{equation}

Using integral $I_1$, one gets the equation
\begin{equation}
\frac{\D \theta}{ \sqrt{k+\sin^2(\theta)}}=\omega \D \tau\, ,
\label{en-bigrot}
\end{equation}
where $k=2 I_1/\omega^2$.  In the variable $x=\sin(\theta)$, equation (\ref{en-bigrot}) produces the standard incomplete elliptic integral and 
hence,
\begin{equation}
\int \frac{\D x}{(1-x^2)(k+x^2)}+\omega \tau=const\, .
\end{equation}
For $k>0$ the integration gives (see 214.00 in \cite{BF71})
\begin{equation}
\frac{1}{\sqrt{1+k}} F \left( \arcsin\left( \sqrt{\frac{(1+k)x^2}{k+x^2}}\right),\frac{1}{\sqrt{1+k}}\right) +\omega \tau=const\, .
\label{x+bigrot}
\end{equation}
At $k<0$ one has (see 214.00 in \cite{BF71})
\begin{equation}
F \left( \arcsin\left( \sqrt{\frac{k+x^2}{(1+k)x^2}}\right),{\sqrt{1+k}}\right)  +\omega \tau=const\, .
\label{x-bigrot}
\end{equation}
The formulae (\ref{x+bigrot}) and  (\ref{x-bigrot}) provide us the third integral $I_3$ for the equation (\ref{chareqn-bigrot}). 

Next, using the integral $I_2$ (\ref{enmom-bigrot}), one gets the equation
\begin{equation}
\D \phi=(I_2 - \omega\log x^2(\tau ))\D \tau
\end{equation}
where $x(\tau)$ is given by (\ref{x+bigrot}) or (\ref{x-bigrot}), namely, $k>0$
\begin{equation}
x^2(\tau)= \frac
{k \, \mathrm{sn}^2 \left(I_3-\omega \tau \sqrt{1+k^2}, \frac{1}{\sqrt{1+k^2}} \right)}
{1+k-\mathrm{sn}^2 \left(I_3-\omega \tau \sqrt{1+k^2}, \frac{1}{\sqrt{1+k^2}} \right)}\, .
\end{equation}
So, the fourth integral is given by ($k>0$)
\begin{equation}
I_4= \phi - \int \left( I_2 - \omega\log  \frac
{k \, \mathrm{sn}^2 \left(I_3-\omega \tau \sqrt{1+k^2}, \frac{1}{\sqrt{1+k^2}} \right)}
{1+k-\mathrm{sn}^2 \left(I_3-\omega \tau \sqrt{1+k^2}, \frac{1}{\sqrt{1+k^2}} \right)}\right) \D \tau\, .
\label{intvar-bigrot}
\end{equation}
The integral hypersurface for the system (\ref{EE-bigrot}) is given by the equation
\begin{equation}
S_i(I_1,I_2,I_3,I_4)=0\, , \qquad i=1,2\, ,
\label{solhodo-bigrot}
\end{equation}
where $S_1$ and $S_2$ are arbitrary functions.

Resolving (\ref{solhodo-bigrot}) w.r.t. $I_3$ and $I_4$, one obtains the hodograph equation 
\begin{equation}
I_3=\Phi_1(I_1,I_2)\, , \qquad I_4=\Phi_2(I_1,I_2)\, ,
\end{equation}
and finally solutions of the system (\ref{EE-bigrot}).

The blowup curve for these solutions is given by the equation
\begin{equation}
\det
\begin{pmatrix}
u \pp{\Phi_1}{I_1}-\pp{I_3}{u} &  &\pp{\Phi_1}{I_2}-\pp{I_3}{v} 
\\ &&  \\
u \pp{\Phi_2}{I_1}-\pp{I_4}{u} &  & \pp{\Phi_2}{I_2}-\pp{I_4}{v} 
\end{pmatrix}=0\, .
\end{equation}

Single-valued solutions of the equation (\ref{EE-bigrot}) are provided by the hodograph equation
\begin{equation}
I_2=\Phi_1(L_1,L_2)\, , \qquad  \mathcal{T}(I_4)=\Phi_2(L_1,L_2)\, ,
\end{equation}
where $ \mathcal{T}(I_4)$ is any periodic function such that $ \mathcal{T}(I_4)= \mathcal{T}(I_4+2\pi)$, for example $ \mathcal{T}(I_4)=\sin(I_4)$.

\section{Rapidly rotating sphere: the case of Coriolis force}
\label{sec-bigrotCor}
In the general case, considered in the previous sections, the role of the centrifugal force, due to the presence of the $\omega^2$ term in 
the equation (\ref{EE-bigrot}) is crucial.

However, in several papers for various reasons the limit $\omega \gg |v|$ has been considered in the presence of only Coriolis force 
(see e.g. \cite{TDBZ14,CG22}). Here we will present the corresponding hodograph equations. In the limit $\omega \gg |v|$ the equations 
(\ref{EE-smallrot}) assume the form
\begin{equation}
\begin{split}
& \pp{u}{t}+ u \pp{u}{\theta}+v \pp{u}{\phi}=2 v \omega \sin(\theta) \cos(\theta) \, , \\
& \pp{v}{t}+ u \pp{v}{\theta}+v \pp{v}{\phi}=-2 \omega u\,  \mathrm{cotan}{(\theta)} \, , \\
\end{split}
\label{EE-Corbigrot}
\end{equation}
and equations of characteristics become
\begin{equation}
\frac{\D \theta}{\D  \tau}=u\, , \qquad
\frac{\D \phi}{\D  \tau}=v\, , \qquad
\frac{\D u}{\D  \tau}=2 v \omega \sin(\theta) \cos(\theta)\, , \qquad
\frac{\D v}{\D  \tau}=-2 \omega u\, \mathrm{cotan}{(\theta)}\, .
\label{chareqn-Corbigrot}
\end{equation}
It is worth to note that  the equation (\ref{EE-Corbigrot}) implies 
\begin{equation}
\frac{\D u^2}{\D t}+\sin^2(\theta) \frac{\D v^2}{\D t}=0\, ,
\end{equation}
and for the characteristic equation
\begin{equation}
\frac{\D u^2}{\D \tau}+\sin^2(\theta) \frac{\D v^2}{\D \tau}=0\, .
\end{equation}
First two integrals for the system (\ref{chareqn-Corbigrot}) can be easily calculated. They are (cfr. formula (\ref{enmom-bigrot})) 
\begin{equation}
I_1=\frac{1}{2}u^2- \omega(v+\omega) \sin^2(\theta)\, , \qquad I_2=v+2 \omega \log\sin(\theta)\, .
\label{enmom-Corbigrot}
\end{equation}
These two integrals are, in fact, certain limits of two particular elements of the ring of integrals for the system (\ref{hodoPDE-Cor}). 
Indeed, let us consider the integral $I^*_1=H-\omega L_3$ where $H$ and $L_3$ where $H$ and $L_3$ are given in (\ref{En-Cor}) 
and (\ref{mom-Cor}).
Rewriting $I^*_1$ in the form
\begin{equation}
I^*_1=\frac{1}{2}u^2-\omega^2 \left(1+\frac{v}{\omega}-\frac{1}{2}\left(\frac{v}{\omega}\right)^2\right) \sin^2(\theta)\, ,
\end{equation}
and dropping in the limit $\frac{|v|}{\omega} \ll 1$ the term with $\left(\frac{v}{\omega}\right)^2$, one obtains the integral $I_1$.
In order to obtain the integral $I_2$ one can proceed as in section \ref{sec-bigrot}. The use of only two integrals $I_1$ and $I_2$   
(\ref{enmom-Corbigrot}) allows us to construct the constraints for the equation (\ref{EE-Corbigrot}) given by the equation
\begin{equation}
S(I_1,I_2)=0\, ,
\label{impl-hodosol-Corbigrot}
\end{equation}
where $S$ is an arbitrary function. Resolving (\ref{impl-hodosol-Corbigrot}) w.r.t. $I_2$, one gets the relation
\begin{equation}
 u^2= 2 \omega (v+\omega) \sin^2(\theta)+2 \Phi(v+2\omega \log\sin(\theta))
\label{hodosol-Corbigrot}
\end{equation}
where $\Phi$ is an arbitrary function.  It is a simple check that the relation (\ref{hodosol-Corbigrot}) represent an admissible constraint for
the system (\ref{EE-Corbigrot}). In this case it is sufficient to solve the second equation  (\ref{EE-Corbigrot}) with $u$ given by the formula
(\ref{hodosol-Corbigrot}). Substituting the obtained $v$ into (\ref{hodosol-Corbigrot}), one gets the function $u(t,\theta,\phi)$.
In the particular case of solutions depending only on $\theta$ one has
\begin{equation}
u(\theta)= \sqrt{2 \Phi(a) + 2 \omega (\omega +a -2 \omega \log \sin(\theta)) \sin^2(\theta)  }\, , 
\qquad v(\theta)=a-2 \omega \log \sin(\theta) \, \qquad a \in \mathbb{R}\, .
\end{equation}

In order to construct general solutions of the system (\ref{EE-Corbigrot}) one needs to find additional functionally independent integrals of 
equation (\ref{chareqn-Corbigrot}). Using the integral $I_2$, one rewrites the expression for the integral $I_1$ in the form
\begin{equation}
\left(\frac{\D \theta}{\D t} \right)^2= 2I_2 +2 \omega (I_1+\omega-\omega \log\sin^2(\theta)) \sin^2(\theta) \, .
\end{equation}
So, one has
\begin{equation}
\frac{\D \theta}{\sqrt{I_2 + \omega (I_1+\omega-\omega \log\sin^2(\theta)) \sin^2(\theta)}}=\sqrt{2} \D \tau\, .
\end{equation}
Introducing the variable $x=\sin(\theta)$, one rewrites the previous equation as
\begin{equation}
\frac{\D x}{\sqrt{(1-x^2)(I_2+\omega(\omega+I_1)x^2-\omega^2 x^2 \log(x^2))}}=\sqrt{2} \D \tau\, ,
\end{equation}
and, consequently, one gets the integral 
\begin{equation}
I_3=\sqrt{2} \tau+\mathcal{G}(\sin^2(\theta),\omega,I_1,I_2)
\label{cq3-Corbigrot}
\end{equation}
where 
\begin{equation}
\mathcal{G}(x,\omega,I_1,I_2)=\int^x \frac{\D \xi}{\sqrt{(1-\xi^2)(I_2+\omega(\omega+I_1)\xi^2-\omega^2 \xi^2 \log(\xi^2))} }\, .
\end{equation}
Substituting (\ref{cq3-Corbigrot}) $\tau$ by $t$, $I_1$ and $I_2$ by the expression (\ref{hodosol-Corbigrot}), one obtains solution
of the equation $S_i$.
In order to find fourth integral one uses the integral $I_2$ which gives
\begin{equation}
\frac{\D \phi}{\D t}=I_2-\omega \log\sin^2(\theta)\, .
\end{equation}
Using the expression for $\sin^2(\theta)$ from (\ref{cq3-Corbigrot}), one obtains
\begin{equation}
\phi=I_2 t-\omega \int^t \log\sin^2(\theta(\xi))\D \xi +const \, ,
\end{equation}
and, consequently, the integral
\begin{equation}
I_4=\phi +\omega \int^t \log\sin^2(\theta(\xi))\D \xi -  (v+2 \omega \log\sin(\theta))t\, .
\end{equation}
Integrals $I_1$, $I_2$, $I_3$, and $I_4$ are functionally independent.  So the integral hypersurface generically is given by the equation
\begin{equation}
S_i(I_1,I_2,I_3,I_4)=0\, , \qquad i=1,2\, ,
\end{equation}
where $S_1$ and $S_2$ are arbitrary functions.
So one gets the hodograph equations
\begin{equation}
I_3=\Phi_1(I_1,I_2)\, , \qquad  I_4=\Phi_2(I_1,I_2)\, ,
\end{equation}
where $\Phi_1$ and $\Phi_2$ are arbitrary functions and, consequently solutions of the equation (\ref{EE-Corbigrot}). The derivatives of this equation
blows-up on the curve
\begin{equation}
\det \begin{pmatrix}
u \pp{\Phi_1}{I_1}-\pp{I_3}{u} && \pp{\Phi_1}{I_2}-\omega \sin^2(\theta)\pp{\Phi_1}{I_1}-\pp{I_3}{v} 
\\ &&  \\
u \pp{\Phi_2}{I_1}-\pp{I_4}{u} && \pp{\Phi_2}{I_2}-\omega \sin^2(\theta)\pp{\Phi_2}{I_1}-\pp{I_4}{v} 
\end{pmatrix}=0\, . 
\end{equation}

Single-valued solutions of the equation (\ref{EE-Corbigrot}) are provided by the hodograph equation
\begin{equation}
I_3=\Phi_1(L_1,L_2)\, , \qquad  \mathcal{T}(I_4)=\Phi_2(L_1,L_2)\, ,
\end{equation}
where $ \mathcal{T}(I_4)$ is any periodic function such that $ \mathcal{T}(I_4)= \mathcal{T}(I_4+2\pi)$.

\section{The reduction $v+\omega=0$}
\label{sec-intvel}
The Euler equation (\ref{EE-Alisei}) admits various constraints. The constraint $v(t,\theta,\phi)=-\omega$ is the simplest one and,
apparently almost trivial.
\subsection{Generic case}
Under this constraint the generic Euler equation (\ref{EE-Alisei}) is reduced to the single equation 
\begin{equation}
 \pp{u}{t}+ u\pp{u}{\theta}-\omega \pp{u}{\phi}=0\, .
 \label{vertHopf-red}
 \end{equation}
The corresponding characteristic equations (\ref{chareqn}) at $v=-\omega$ have three obvious integrals
\begin{equation}
I_1=u\, , \qquad I_2=\theta- ut\, , \qquad I_1=\phi+\omega t\,  .
\end{equation}
Integral hypersurface is given by the equation
\begin{equation}
S(u,\theta- ut,\phi+\omega t)=0\, ,
\label{Corred-hyper}
\end{equation}
which provides us with the hodograph equation
\begin{equation}
\theta-ut=\Phi(u, \phi+\omega t)\, ,
\end{equation}
where $\Phi$ is an arbitrary function inverse to the initial data $u(\theta,\phi,t=0)$.

The derivatives of $u$ of the solutions of the equation blow-up on the curve $\pp{\phi}{u}+t=0$. Equation (\ref{vertHopf-red}) is connected
$\pp{\tilde{u}}{t}+\tilde{u} \pp{\tilde{u}}{\theta}=0$ by the transformation (\ref{map-rnr}) with $\tilde{v}=0$.

\subsection{Coriolis force case}
Less trivial situation occurs for the Euler equation which contains only the Coriolis force, i.e. for the system (\ref{EE-smallrot}). 
In this case, under the constraint $v=-\omega$ one gets the equation
\begin{equation}
\pp{u}{t}+u \pp{u}{\theta}-\omega  \pp{u}{\phi}=-\omega^2\sin(\theta)\cos(\theta)\, .
\label{EE-Corred}
\end{equation}
Characteristic equations (\ref{chareqn-Cor}) become
\begin{equation}
\frac{\D t}{\D \tau}=1\, , \qquad
\frac{\D \theta}{\D \tau}=u\, , \qquad
\frac{\D \phi}{\D \tau}=-\omega\, , \qquad
\frac{\D u}{\D \tau}=-\omega^2\sin(\theta)\cos(\theta)\, ,  \qquad
\frac{\D v}{\D \tau}=0\, , 
\label{char-redint}
\end{equation}
or, equivalently,
\begin{equation}
\frac{\D^2 \theta}{\D \tau^2}+\omega^2\sin(\theta)\cos(\theta)=0\, .
\end{equation}
This equation for $\theta$ is the equation of motion for the mathematical pendulum which is a classical textbook problem (see e.g. \cite{L-I}).
Equation (\ref{char-redint}) have three integrals ($L_3=0$) 
\begin{equation}
H=\frac{1}{2} u^2+\frac{1}{2} \omega^2 \sin^2(\theta)\, , \qquad
{P}=-\sqrt{2H}\, t + F\left( \theta, \frac{\omega}{\sqrt{2H}}\right)\, , \qquad
N=\phi+\omega t\, .
\end{equation}
where $F(\theta,k)$ is the incomplete elliptic integral of the first kind.
So the integral hypersurface is given by 
\begin{equation}
S(H,{P},\phi+\omega t)=0\, .
\end{equation}
Resolving  this equation w.r.t. $P$, one obtains the hodograph equation 
\begin{equation}
-\sqrt{ u^2+ \omega^2 \sin^2(\theta)}\, t + F\left(\theta, \frac{\omega}{\sqrt{u^2+ \omega^2 \sin^2(\theta)}} \right)=
\Phi(u^2+ \omega^2 \sin^2(\theta),\phi+\omega t)\, ,
\label{vertredCor-hodo}
\end{equation}
where $\Phi$ is an arbitrary function.

Differentiating  (\ref{vertredCor-hodo}) w.r.t. $t$, $\theta$, and $\phi$, one gets 
\begin{equation}
\begin{split}
-\sqrt{u^2+ \omega^2 \sin^2(\theta)}-\omega \pp{\Phi}{\phi}&=M \pp{u}{t}\, , \\
\sqrt{u^2+ \omega^2 \sin^2(\theta)}&=M \left( u \pp{u}{\theta}+\omega^2 \sin(\theta)\cos(\theta) \right)\, \\
- \pp{\Phi}{\phi}&=M \pp{u}{\phi}\, ,  
\end{split}
\label{hododer-Corred}
\end{equation}
where
\begin{equation}
M= \left({u^2+ \omega^2 \sin^2(\theta)}\right)^{-1/2} \, ut + u \omega \Big({u^2+ \omega^2 \sin^2(\theta)}\Big)^{-3/2} \, \pp{F}{\xi} + 2 u \Phi' \, ,
\end{equation}
and $\pp{F}{\xi}$ indicates $\pp{F(\theta,\xi)}{\xi}$ while $\Phi'$ indicates the derivative of $\Phi$ w.r.t. the first argument.
Combining the relations (\ref{hododer-Corred}), one obtains
\begin{equation}
M \left( \pp{u}{t}+u \pp{u}{\theta}-\omega  \pp{u}{\phi}+\omega^2\sin(\theta)\cos(\theta)\right)=0\, .
\end{equation}
So, if $M \neq 0$ the solution $u(t,\theta,\phi)$ of the hodograph equation (\ref{vertredCor-hodo}) obeys the equation (\ref{EE-Corred}). 
On the case $M=0$, the derivatives of $u$ blow-up.

So, the hodograph equation (\ref{vertredCor-hodo}) gives us general solution of the equation (\ref{EE-Corred}) in the form
\begin{equation}
u=\sqrt{\zeta(t,\theta,\phi)-\omega^2 \sin^2(\theta)}\, , \qquad
\end{equation}
where the function $\zeta=\zeta(t,\theta,\phi)$ is a solution of the trascendental equation
\begin{equation}
-\zeta \, t + F \left( \theta,\frac{\omega}{\zeta} \right)- \tilde{\Phi}(\zeta,\phi+\omega t)=0\, ,
\end{equation}
and  $\tilde{\Phi}$ an arbitrary function.

\begin{figure}[h!]
\begin{center}
\includegraphics[width=.4 \textwidth]{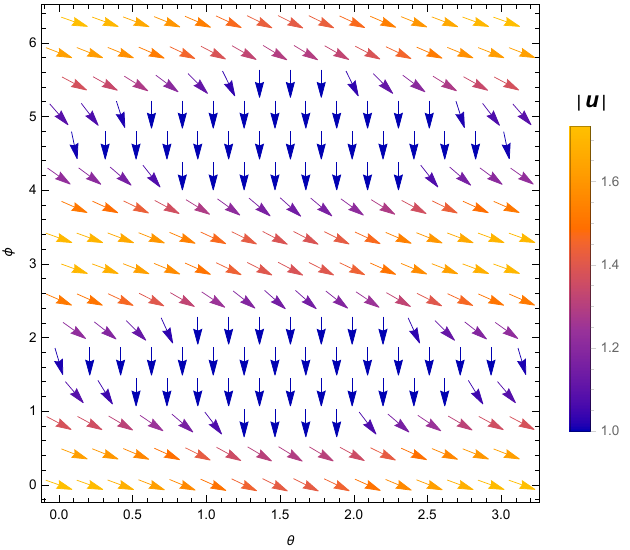}
\caption{The stationary solution (\ref{solCorsol-stat}) when $\tilde{\Phi}(x)=\cos(2x)$ is $\uu=\Big( 1 + \cos (2(\phi + \omega t)) - \omega^2 \sin^2(\theta) ,-\omega \Big)$. It is presented at $\omega=1$, $t=0$. }
\label{Stat-Cor-fig}
\end{center}
\end{figure}

It is worth to note that the different choices of the function $S$ give other particular solution of the equation (\ref{EE-Corred}). For instance, with
$S=\Psi(H,N)$, one gets the solution
\begin{equation}
u=\pm \sqrt{\tilde{\Phi}(\phi+\omega t) -\omega^2 \sin^2(\theta)} \, ,
\label{solCorsol-stat}
\end{equation}
where $\tilde{\Phi}$ is an arbitrary function. 
The plot of a simple solution in this family is  in figure \ref{Stat-Cor-fig} .

We note also some equivalent forms of the equation (\ref{EE-Corred}). First, rewritten in the conservation law form
\begin{equation}
\left( \pp{}{t} -\omega \pp{}{\phi} \right) u+\pp{}{\theta } \left( \frac{1}{2}u^2+ \frac{1}{2}\omega^2 \sin^2(\theta) \right) =0\, ,
\end{equation}
it implies the Hamilton-Jacobi type equation for the potential $\psi$ $\left(u=\pp{\psi}{\theta}\right)$:
\begin{equation}
\left( \pp{}{t} -\omega \pp{}{\phi} \right) \psi+  \frac{1}{2} \left( \pp{\psi}{\theta} \right)^2 + \frac{1}{2}  \omega^2 \sin^2(\theta) =0\, .
\end{equation}
Second, the change of variables $\tilde{t}=t$, $\tilde{\phi}=\phi +\omega t$, $\tilde{u}=u(\tilde{t},\theta,\tilde{\phi}-\omega \tilde{t})$  
changes the equation (\ref{EE-Corred}) into the following equation
\begin{equation}
 \pp{\tilde{u}}{\tilde{t}}+ \tilde{u}\pp{\tilde{u}}{\theta}+\omega^2 \sin(\theta) \cos(\theta)=0\, .
 \end{equation}
Equations of such type has been discussed in several papers (see e.g. \cite{CF03}). 

\subsection{Equation for elliptic modulus} 
The form of the hodograph equation (\ref{vertredCor-hodo})  suggests also the introduction of new dependent variables
\begin{equation}
k=\frac{\omega}{\sqrt{u^2+\omega^2 \sin^2(\theta)}}\, ,
\end{equation}
that is the elliptic modulus for the function $F(\theta,k)$, or $k=\omega/\sqrt{2H}$. The corresponding equation is
\begin{equation}
\pp{k}{t}
+\frac{\omega \sqrt{1-k^2 \sin^2(\theta)}}{ k} \pp{k}{\theta} 
- \omega \pp{k}{\phi}=0\, .
\label{EE-Corred-ell}
\end{equation}
The hodograph equation for this equation is given by 
\begin{equation}
-\omega t + k F(\theta,k)=\tilde{\Phi}(k,\phi+\omega t)\, ,
\label{hodo-Corred-ell}
\end{equation}
where $\tilde{\Phi}$ is an arbitrary function. 
It is a standard check that solutions of the equation (\ref{hodo-Corred-ell}) are generically solutions of equation (\ref{EE-Corred-ell}). 
The blowup curve is given by
\begin{equation}
F(\theta,k)+ k \pp{F(\theta,k)}{k}-\pp{\tilde{\Phi}(k,\phi+\omega t)}{k}=0\, .
\end{equation}
It is noted that the formulae (\ref{EE-Corred-ell}) and (\ref{hodo-Corred-ell}) can be interpreted as equations describing  a class of deformations 
of the normal elliptic integral of the first kind
\begin{equation}
F(\theta,k)=\int_0^\theta \frac{\D \xi}{\sqrt{1-k^2 \sin^2(\xi)}}\, .
\end{equation}
Indeed, let the elliptic modulus $k$ is deformed, i.e. $k=k(t, \theta)$. Then the equation 
(equation (\ref{EE-Corred-ell}) with $\pp{k}{\phi}=0$ and $\omega=1$) 
\begin{equation}
\pp{k}{t}
+\frac{ \sqrt{1-k^2 \sin^2(\theta)}}{ k} \pp{k}{\theta} =0\, ,
\end{equation}
describes deformations of $k$ such that
\begin{equation}
k F(\theta,k)=t+\Phi(k)\, ,
\end{equation}
where $\Phi(k)$ is an arbitrary function. Possible geometric meaning of such deformations will be discussed elsewhere.


\section{On the Euler equation for physical velocities}
\label{sec-physvel}
The velocities $u$ and $v$ in the Euler equation (\ref{EE-Alisei}) are derivatives of the local coordinates $\theta$ and $\phi$ w.r.t. time: 
$u=\pp{\theta}{t}$ and $v=\pp{\phi}{t}$. In these variables the Euler equations for nonrotating sphere ($\omega=0$) is invariant under the general
coordinate transformation (see e.g. \cite{Gat04}).
The physical velocities  are vectors tangent to the sphere surface with the components
\begin{equation}
\tilde{u}=u\, , \qquad \tilde{v}=v \sin(\theta)\, .
\label{physvelmap}
\end{equation}
 In terms of $\tilde{u}$ and $\tilde{v}$ the Euler equations (\ref{EE-Alisei}) assumes the form
 \begin{equation}
 \begin{split}
\pp{\tilde{u}}{t}+\tilde{u} \pp{\tilde{u}}{\theta} + \frac{\tilde{v}}{\sin(\theta)} \pp{\tilde{u}}{\phi}
&=(\tilde{v}+\omega \sin(\theta))^2 \, \mathrm{cotan}(\theta) \, ,  \\ 
\pp{\tilde{v}}{t}+\tilde{u} \pp{\tilde{v}}{\theta} + \frac{\tilde{v}}{\sin(\theta)} \pp{\tilde{v}}{\phi}
&=-\tilde{u}(\tilde{v}+2\omega \sin(\theta)) \, \mathrm{cotan}(\theta)\,  \\
  \end{split}
  \label{EE-Alisei-phys}
 \end{equation}
and has been discussed in a number of papers (see e.g. \cite{Gat04,TDBZ14}).

In the generic case the solutions of the system (\ref{EE-Alisei-phys}) are obtained from those presented in sections (\ref{sec-EEgen}) 
and (\ref{sec-genhodo}) by the simple transformation (\ref{physvelmap}). The same occurs for the limit of small $\omega$. 
Indeed, for small $\omega \ll |v|$, equation  (\ref{EE-Alisei-phys}) are approximated by the system
 \begin{equation}
 \begin{split}
\pp{\tilde{u}}{t}+\tilde{u} \pp{\tilde{u}}{\theta} + \frac{\tilde{v}}{\sin(\theta)} \pp{\tilde{u}}{\phi}
&=\tilde{u}(\tilde{v}+2\omega \sin(\theta)) \, \mathrm{cotan}(\theta) \, ,  \\ 
\pp{\tilde{v}}{t}+\tilde{u} \pp{\tilde{v}}{\theta} + \frac{\tilde{v}}{\sin(\theta)} \pp{\tilde{v}}{\phi}
&=-\tilde{u}(\tilde{v}+2\omega \sin(\theta)) \, \mathrm{cotan}(\theta)\, , \\
  \end{split}
  \label{EE-smallrot-phys}
 \end{equation}
Equations of characteristics for (\ref{EE-smallrot-phys}) are given by
\begin{equation}
\frac{\D t}{\D \tau}=1\, , \quad
\frac{\D \theta}{\D \tau}=\tilde{u}\, , \quad
\frac{\D \phi}{\D \tau}=\frac{\tilde{v}}{\sin(\theta)}\, , \quad
\frac{\D \tilde{u}}{\D \tau}=\tilde{u}(\tilde{v}+2\omega \sin(\theta)) \, \mathrm{cotan}(\theta)\, , \quad
\frac{\D \tilde{v}}{\D \tau}=-\tilde{v}(\tilde{v}+2\omega \sin(\theta)) \, \mathrm{cotan}(\theta)\, .
  \label{chareqn-Cor-phys}
\end{equation}
It is an easy check that the equations (\ref{EE-smallrot}), (\ref{chareqn-Cor}) and (\ref{EE-smallrot-phys}), (\ref{chareqn-Cor-phys}), 
and, consequently, their solutions, are connected by the transformation (\ref{physvelmap}).

For rapidly rotating sphere (large $\omega \gg |v|$) the situation is apparently different. Arguing in the same way, as for the equations 
(\ref{EE-bigrot}), i.e. leaving in the r.h.s.  of equation (\ref{EE-Alisei-phys}) only the terms with highest order in $\omega/|v|$, 
one gets the equations
 \begin{equation}
 \begin{split}
\pp{\tilde{u}}{t}+\tilde{u} \pp{\tilde{u}}{\theta} + \frac{\tilde{v}}{\sin(\theta)} \pp{\tilde{u}}{\phi}
&= \omega^2 \sin(\theta) \cos(\theta) \, ,  \\ 
\pp{\tilde{v}}{t}+\tilde{u} \pp{\tilde{v}}{\theta} + \frac{\tilde{v}}{\sin(\theta)} \pp{\tilde{v}}{\phi}
&=-2\tilde{u}\omega \cos(\theta) \, . \\
  \end{split}
  \label{EE-bigrot-phys}
 \end{equation}
The corresponding equations of characteristics are
\begin{equation}
\frac{\D t}{\D \tau}=1\, , \quad
\frac{\D \theta}{\D \tau}=\tilde{u}\, , \quad
\frac{\D \phi}{\D \tau}=\frac{\tilde{v}}{\sin(\theta)}\, , \quad
\frac{\D \tilde{u}}{\D \tau}=\omega^2 \sin(\theta) \cos(\theta)\, , \quad
\frac{\D \tilde{v}}{\D \tau}=-2\tilde{u}\omega \cos(\theta)\, .
  \label{chareqn-bigrot-phys}
\end{equation}
It is a direct check that the equations for $v$ and $\tilde{v}$ and, hence, the whole system (\ref{EE-bigrot}) , (\ref{chareqn-bigrot}) 
and (\ref{EE-bigrot-phys}) , (\ref{chareqn-bigrot-phys})   are not related by the map (\ref{physvelmap}).
So the na\"ive limit for large $\omega \gg |v|$ of the systems (\ref{EE-Alisei}) and (\ref{EE-Alisei-phys}) are formally different: they are interconnected 
by the transformation (\ref{physvelmap}) modulo the small term $uv \ll 2 \omega u$.

Let us construct the hodograph equations for the system (\ref{EE-bigrot-phys}). 
The formulae  (\ref{chareqn-bigrot-phys}) immediatly gives us two integrals
\begin{equation}
I_1= \frac{1}{2} \tilde{u}^2-\frac{1}{2} \omega^2  \sin^2(\theta)\, , \qquad I_2=\tilde{v}+ 2 \omega  \sin(\theta)\, . 
\label{enmom-bigrot-phys}
\end{equation}
Using the integral $I_1$, one gets equation (\ref{en-bigrot}) and formulae (\ref{x+bigrot}), (\ref{x-bigrot}) and, consequently the integral $I_3$, 
the same as in section \ref{sec-bigrot}. 

Then, using the integral $I_2$ in (\ref{enmom-bigrot-phys}), one obtains the equation
\begin{equation}
\D \phi + 2 \omega \D \tau = \frac{I_2}{x(\tau)}\D \tau\, ,
\end{equation}
where $x(\tau)$ is given by (\ref{intvar-bigrot}).

Consequently, one gets the fourth integral
\begin{equation}
I_4= \phi +2\omega \tau -\frac{I_2}{2 k} \int^\tau  \left(   \frac{\sqrt{k + \mathrm{cn}^2\left( I_3 -\sqrt{1+k} \omega \xi , \frac{1}{\sqrt{1+k}}\right)}}{\mathrm{sn}\left( I_3 -\sqrt{1+k} \omega \xi , \frac{1}{\sqrt{1+k}}\right)  } \right)  \D \xi \, .
\end{equation}
So choosing the integral hypersurface by the equation
\begin{equation}
S_i(I_1,I_2,I_3,I_4)=0, \qquad i=1,2\, ,
\end{equation}
where $S_1$ and $S_2$ are arbitrary functions, one gets the hodograph equations for the system (\ref{EE-bigrot-phys}).  For instance
\begin{equation}
\tilde{u}^2 -  \omega^2 \sin^2(\theta)=\Phi_1(I_3,I_4)\, , \qquad \tilde{v}+ 2 \omega \sin(\theta)=\Phi_2(I_3,I_4)\, ,
\label{hodo-bigrot-phys}
\end{equation}
where $\Phi_1$ and $\Phi_2$ are arbitrary functions.
Solutions of the system (\ref{hodo-bigrot-phys}) are generically solutions of the systems (\ref{EE-bigrot-phys}). Derivatives of these solutions
blow-up on the curve
\begin{equation}
\det
\begin{pmatrix}
 \pp{\Phi_1}{I_3} \pp{I_3}{\tilde{u}}+\pp{\Phi_1}{I_4} \pp{I_4}{\tilde{u}}-2 \tilde{u}&&  \pp{\Phi_1}{I_3} \pp{I_3}{\tilde{v}}+\pp{\Phi_1}{I_4} \pp{I_4}{\tilde{v}}
 \\&&\\
  \pp{\Phi_2}{I_3} \pp{I_3}{\tilde{u}}+\pp{\Phi_2}{I_4} \pp{I_4}{\tilde{u}} &&  \pp{\Phi_2}{I_3} \pp{I_3}{\tilde{v}}+\pp{\Phi_2}{I_4} \pp{I_4}{\tilde{v}}-1
\end{pmatrix}
=0\, .
\end{equation}
Particular properties of the specific classes of exact solutions of the Euler equations presented above will be analysed elsewhere.

Single-valued solutions of the system (\ref{EE-bigrot-phys}) are obtained with  the use of the hodograph equation
\begin{equation}
\tilde{u}-\omega^2 \sin^2(\theta)=\Phi_1(I_3,\mathcal{T}(I_4))\, , \qquad
\tilde{v}+2\omega \sin(\theta)=\Phi_2(I_3,\mathcal{T}(I_4))\, , \qquad
\end{equation}
where $ \mathcal{T}(I_4)$ is a generic periodic function such that $ \mathcal{T}(I_4)= \mathcal{T}(I_4+2\pi)$.

Solutions of the Euler equation (\ref{EE-Alisei}) without singularities and with some particular properties relevant for applications in physics 
will be discussed elsewhere.

%

\subsubsection*{Acknowledgments}
The authors are grateful to L. Fatibene  and A. A. Veselov  for useful discussions.
This project has received funding from the European Union's Horizon 2020 research and innovation programme under the Marie Sk{\l}odowska-Curie grant no 778010 {\em IPaDEGAN}  and by the PRIN 2022TEB52W-PE1 Project ``The charm of integrability: from nonlinear waves to random matrices". We also gratefully acknowledge the auspices of the GNFM Section of INdAM, under which part of this work was carried out, and the financial support of the project MMNLP (Mathematical Methods in Non Linear Physics) of the INFN.


\appendix 
\section{Solutions with constant $\mathbf{L_3}$} 
\label{app-constantL3}
Simple solutions of the hodograph equations (\ref{simple-hodo}) are obtained for the choice $\Psi_1=0$ and $\Psi_2=L_3=$const. Indeed
in this case
\begin{equation}
\begin{split}
&t+\sigma \frac{1}{\sqrt{u^2+\sin^2(\theta) (v+\omega)^2}} Q(u,v,\theta)=0\, , \\
&v=\frac{\Phi_2}{\sin^2(\theta)}-\omega\, .
\end{split}
\label{simple-hodo-app}
\end{equation}
In the particular case when $u$ does not depends on $t$ the quantity
\begin{equation}
u^2+\frac{\Phi_2^2}{\sin^2(\theta)}=A=\text{const.}
\end{equation}
is an integral of the equation (\ref{simple-hodo-app}). 

So one has the solution
\begin{equation}
u_\pm=\pm \sqrt{A- \frac{\omega^2}{\sin^2(\theta)}}\, , \qquad v=\omega \, \mathrm{cotan}^2(\theta)\, .
\label{sol-om-stat}
\end{equation}
In the formal limit $\omega \to 0$ the solution $u_\pm, v$ becomes
\begin{equation}
u_\pm\vert_{\omega=0}= \pm \sqrt{A}\, , \qquad v_\pm\vert_{\omega=0}= 0\, .
\end{equation}
Note that another solution 
\begin{equation}
u_\pm = \pm  \sqrt{A}\, , \qquad v_\pm= -\omega \, ,
\label{trivconstsol}
\end{equation}
has the same limit $u=\pm \sqrt{A}$, $v=0$ at $\omega=0$.

It is worth to emphasize that the transformation (\ref{invmap-rnr}) converts the solution (\ref{sol-om-stat}) into the solution
\begin{equation}
\tilde{u}=\sqrt{A-\frac{\omega^2}{\sin^2(\theta)}}\, , \qquad \tilde{v}=\frac{\omega}{\sin^2(\theta)}
\end{equation}
of the equation (\ref{EE-Sphere}) while the transformation of the solution (\ref{trivconstsol}) gives
\begin{equation}
u_\pm\vert_{\omega=0}= \pm  \sqrt{A}\, , \qquad v_\pm\vert_{\omega=0}= 0\, .
\end{equation}
This fact illustrates the difference between the transformation (\ref{invmap-rnr}) and the formal limit $\omega \to 0$ for the solutions
of the equation (\ref{EE-Alisei}).

\section{Lagrangian  of the characteristics of the system (\ref{EE-smallrot}) }
\label{app-eqstart}
We report here a Lagrangian structure of the characteristics of (\ref{EE-smallrot}): they 
 are equivalent to the particle motion on a sphere subject to Coriolis force and this coincide with the Lagrangian interpretation of the  fluid motion.
We report here the equation of characteristics (\ref{char-Newton})
 \begin{equation}
 \begin{split}
 \frac{\D^2 \theta}{ \D t^2}&=\sin \theta \cos \theta \left( \left( \frac{\D \phi}{\D t}+\omega\right)^2-\omega^2\right)\qquad \, ,\\
 \frac{\D^2 \phi}{ \D t^2}&=-2 \frac{ \cos \theta}{\sin \theta} \frac{\D \theta}{\D t} \left( \frac{\D \phi}{\D t}+\omega\right) \qquad \, .
 \end{split}
 \end{equation}
  Such equations are Lagrangian w.r.t. the Lagrangian
  \begin{equation}
  \begin{split}
  \mathcal{L}=&
  \frac{1}{2}\left( \dot{\theta} ^2 + \sin^2 (\theta) \dot{\phi}^2 \right)+ \omega  \sin^2 (\theta)  \dot{\phi} \, \\
  =& \frac{1}{2} \left(\dot{\theta}^2 +\sin^2 (\theta)\left(   \dot{\phi}  +\omega \right) ^2\right) - \frac{1}{2}\omega^2  \sin^2 (\theta)  \, ,
  \end{split}
  \label{Lagr-Cor}
  \end{equation}
  where $\dot{f}= \frac{\D f}{\D t}$.
The conjugate momenta are
\begin{equation}
p_\theta=\pp{\mathcal{L}}{ \dot{\theta}}=\dot{\theta}\, , \qquad p_\phi=\pp{\mathcal{L}}{ \dot{\phi}}= \sin^2(\theta) \left( \dot{\phi} +\omega\right)\, .
\end{equation}
The Jacobi integral associated to the Lagrangian is given by
\begin{equation}
E=p_\theta \dot{\theta}+p_\phi \dot{\phi}-\mathcal{L}= \frac{1}{2} \left(\dot{\theta}^2 +\sin^2 (\theta)  \dot{\phi}^2\right)  \, ,
\end{equation}
which coincide with the energy of a particle on a nonrotating sphere ($\omega=0$).
The Hamiltonian of the system is
\begin{equation}
\mathcal{H}=\frac{p_\theta^2}{2}+\frac{p_\phi^2}{2 \sin^2(\theta)} -\omega p_\phi+\frac{1}{2}\omega^2 \sin^2{\theta}\, .
\end{equation}
The Lagrangian (\ref{Lagr-Cor}) is also the Lagrangian of a particle with unit charge and mass on a sphere in the constant magnetic field 
$\mathbf{B}=(0,0,2 \omega)$.

%
%
%
%

\end{document}